\documentclass[prd, preprint, showpacs, showkeys, nofootinbib]{revtex4}    
\usepackage{amssymb}
\usepackage{amsmath}
\usepackage{graphicx}
\usepackage[dvips]{epsfig}
\usepackage{hyperref}
\usepackage[usenames,dvipsnames]{color}
\hypersetup{
    colorlinks=true,         
    linkcolor=blue,          
    citecolor=red,        
    urlcolor=Violet             
}

\newcommand{\dto}[2]{\ensuremath{#1 =1,\hdots,#2 }}
\newcommand{\eq}[1]{Eq.~(\ref{eq:#1})}

\newcommand{\scn}[1]{Sec.~(\ref{sec:#1})}

\newcommand{\abs}[1]{\ensuremath{\left| #1 \right|}}

\newcommand{\diby}[2]{\ensuremath{\frac{\partial #1}{\partial #2}}}
\newcommand{\equa}[1]{\begin{equation} #1 \end{equation}}
\newcommand{\pb}[2]{\ensuremath{\lf\{#1,#2 \rt\}}}

\def\lf {\ensuremath{\left}}
\def\rt {\ensuremath{\right}}
\def\ra {\ensuremath{\rightarrow}}

\def\dotq {\ensuremath{\dot{q}}}
\def\dotp {\ensuremath{\dot{p}}}
\def\vecq {\ensuremath{\vec{q}}}
\def\vecp {\ensuremath{\vec{p}}}
\def\kpnmt {\ensuremath{\tilde{k}_{\text{PNM}}}}
\def\kpnm {\ensuremath{k_{\text{PNM}}}}
\def\kpnmof {\ensuremath{k_{\text{PNM}}(q^{\prime\prime\alpha},q^{\prime\alpha})}}
\def\kj {\ensuremath{k_{\text{JBB}}}}

\def\leps {\ensuremath{\mathcal{E}}}
\def\fp {\ensuremath{[\text{FP}]}}

\begin{document}

\title{Jacobi's Principle and the Disappearance of Time}
\date{\today}
\author{Sean Gryb}
\affiliation{Perimeter Institute for Theoretical Physics\\Waterloo, Ontario N2L 2Y5, Canada}
\affiliation{Department of Physics and Astronomy, University of Waterloo\\Waterloo, Ontario N2L 3G1, Canada}
\email{sgryb@perimeterinstitute.ca}

\begin{abstract}
	Jacobi's action principle is known to lead to a problem of time. For example, the timelessness of the Wheeler-DeWitt equation can be seen as resulting from using Jacobi's principle to define the dynamics of 3-geometries through superspace. In addition, using Jacobi's principle for non-relativistic particles is equivalent classically to Newton's theory but leads to a time-independent Schr\"odinger equation upon Dirac quantization. In this paper, we study the mechanism for the disappearance of time as a result of using Jacobi's principle in these simple particle models. We find that the path integral quantization very clearly elucidates the physical mechanism for the timeless of the quantum theory as well as the emergence of duration at the classical level. Physically, this is the result of a superposition of clocks which occurs in the quantum theory due to a sum over all histories. Mathematically, the timelessness is related to how the gauge fixing functions impose the boundary conditions in the path integral.
\end{abstract}

\maketitle
\tableofcontents


\section{Introduction}\label{sec:intro}

\subsection{Background}

``It is utterly beyond our power to measure the changes of things by time. Quite the contrary, time is an abstraction, at which we arrive by means of the changes of things...''

In this beautiful quote by Mach from \emph{The Mechanics} \cite{mach:mechanics} dated 1883, he lays out what has been called his \emph{second} principle \cite{Mittelstaedt:machs_2nd}: that time should be a measure of the changes of things. Though the argument seems simple, elegant, and epistemologically sound, it took nearly 100 years before Barbour and Bertotti where able to formulate this principle into a mathematically rigorous theory of time in 1982 \cite{barbourbertotti:mach}\footnote{For a crystal clear account of how Jacobi's principle can be thought as a way of implementing Mach's ideas about time, see \cite{Barbour:nature_of_time}.}. The reason for this delay can not be attributed to technical complications, since the mathematics have been well understood since Jacobi, but rather to conceptual confusion surrounding how Mach's principles are implemented in General Relativity. Though it was clear that Einstein was heavily influenced by Mach's ideas, general covariance proved to be misleading as a way of implementing relationalism. However, through the papers of Barbour and collaborators \cite{barbourbertotti:mach,barbour_el_al:physical_dof,barbour_el_al:rel_wo_rel,barbour_el_al:scale_inv_gravity}, we now have a clear picture of how Mach's ideas can be used to derive General Relativity (GR). From this work, we know that, in regards to time, Mach's second principle can be implemented in GR by using Jacobi's principle to determine the classical dynamics of the system. The consequences of using such a definition of time in the quantum theory lead to a problem of time and are the main concern of this paper. Specifically, we find that, in the path integral description, each path represents a different relational clock. Thus, a sum over all paths leads to a kind of \emph{superposition of clocks} (the precise definition of this will be given in \scn{time}) resulting in a time \emph{independent} theory.

Though there are many different problems of time and many different ways of stating each \cite{kuchar:prob_of_time,kuchar:time_int_qu_gr}, a simple way of stating the most obvious aspect of the problem of time is to note that the Wheeler-DeWitt (WDW) equation \cite{deWitt:equation}
\equa{
  \lf[G^{abcd}(\hat{g}^3) \hat{\pi}_{ab} \hat{\pi}_{cd} + \lf( R^3(\hat{g}^3) - \Lambda \rt) \rt] \Psi[g^3]= 0,
}
in a configuration basis, depends only on the 3-metric $g^3$ and variations with respect to it. Thus, the formal wavefunctional $\Psi[g^3]$ depends only on the configuration space variables $g^3$, and is completely independent of any variable one could interpret as time. As a result, solutions to the WDW equation are stationary states. This fact leads to difficulties, for example, in forming an inner product under which $\Psi$ evolves unitarily and with which one can define a clear notion of probability.

First encountered in the context of GR, this old but surprising result is characteristic of \emph{any} theory that uses Jacobi's principle for determining its classical dynamics. For example, if one uses Jacobi's principle in non-relativistic particle mechanics, one finds (see \scn{ham_jbb}) that the analogue of the WDW equation is the time-independent Schr\"{o}dinger equation (\eq{tise}). As we will see, the result that one gets a quadratic scalar constraint on the wavefunction whose solutions are stationary state comes from using Jacobi's principle to implement Mach's second principle in order to define time in a relational way. As a result, one can study, as we will do here, this aspect of the problem of time in GR by considering the much simpler case of non-relativistic particles moving through a space dependent potential using Jacobi's principle. We will call this theory Jacobi-Barbour-Bertotti (JBB) theory since it was first written down by Jacobi but was given an interpretation in terms of Mach's second principle by Barbour and Bertotti (BB)\footnote{It should be noted that this is different from what is usually referred to as BB theory \cite{barbourbertotti:mach,barbour:eot,barbour:scale_inv_particles,barbour:timelessness,barbour:timelessness2,gergely:geometry_BB1,gergely:geometry_BB2,anderson:triangleland_old} since we are not considering the spatial symmetries which produce linear constraints analogous to the 3-diffeomorphisms of GR. It can be checked \cite{sg:mach_time} that the spacial symmetries add nothing to the discussions regarding time.}. Because the configuration space variables are just simple functions, and not tensor fields on a 3-manifold as is the case in GR, we will not be able to reproduce certain aspects of GR such as the infinity of scalar constraints which lead to Wheeler's \emph{many-fingered time} \cite{bsw:bsw_action}. However, these toy models provide a simple means for studying the global aspects of the disappearance of time and can be applied, for example, directly to symmetry reduced models of GR, like mini-superspace, which are of primary importance in cosmological applications, or any model of gravity with a fixed lapse, such as Ho\v{r}ava's theory \cite{Horava:gravity1}.

\subsection{Summary of Results}

In this paper we will concern ourselves with two main tasks: 1) the physical mechanism, from the point of view of the path integral, for the disappearance of time in the quantum theory, and 2) the conditions underwhich one can recover a notion a time.

Our motivations stem from the following puzzle: as we will see, the canonical quantization of JBB theory leads to the time-independent Schr\"{o}dinger equation whose solutions are stationary states. However, the classical equations of motion give Newton's laws with a specific definition of the Newtonian time in terms of the classical trajectories of the particles in the system (\eq{timej}). Aside from the obvious difficulties associated with defining an inner product for the Hilbert space, it is even less clear why one should expect to obtain a classical limit with a specific notion of time, completely equivalent to Newton's, from a quantum theory which is completely time independent. How could Newton's time be hidden in the time independent Schr\"{o}dinger equation? Alternatively, why is it that, in Newton's theory, time is perfectly well described off-shell while in Jacobi's theory, which is classically equivalent, time is only meaningful on-shell? What has happened to time?

In contrast, the path integral for JBB theory is easy to write down, at least at the formal level. We will find that the analogue of the Newtonian time can be written down in a straightforward way for this path integral. However, in the full quantum theory, this quantity contributes to each term in the sum over all histories leading to a superposition of all possible clocks: the end result being a timeless theory. In the stationary phase approximation, the path integral is dominated by the classical trajectory picking out a unique clock. We will find that this clock does indeed read out a time that is equivalent to that used in Newtonian mechanics. This gives a consistent picture for how a time dependent classical theory could be the limiting form of a time independent quantum theory. Furthermore, this picture also tells us how to obtain a unique notion of duration in the quantum theory. For quadratic potentials, the stationary phase approximation is exact and the Newtonian time along the classical trajectory servers as a unique notion of duration even in the full quantum theory. Taking a close look at the path integral thus provides us with a unique notion of time off-shell (for certain potentials) and intuition as to why this notion of time breaks down in general. This intuition may be invaluable in suggesting new ways inwhich duration may be defined in a time independent quantum theory.

\subsection{Earlier Work}

Admittedly, much has already been written in regards to the problem of time in quantum gravity. Many of the preliminary technical results can be found elsewhere in the literature in a slightly different from. In particular, Lanczos' book \cite{lanczos:mechanics} gives a nice account of Jacobi's principle and Barbour and Bertotti's paper \cite{barbourbertotti:mach} describes the connection to timelessness. The path integral of JBB theory has been explored in \cite{hartlekuchar:pnm} and the BRST quantization in \cite{brown:gr_time}. The path integrals for mini-superspace models (which are very similar to JBB theory) as well as a parameterized form of Newtonian mechanics, which we will be considering and will call Parameterized Newtonian Mechanics (PNM), are treated extensively in \cite{halliwell:mini_ss_PI} and \cite{kiefer:PI_qu_csmlgy}. However, these papers do not address the key results of this paper which are those which concern the emergence of time. The results of \scn{time} show that time does indeed emerge when the stationary phase approximation is exact\footnote{This goes further than the classical limit since certain potentials, like the harmonic oscillator, are exactly described by the stationary phase approximation.} without having to modify the quantization procedure or make an additional ansatz. Furthermore, they illustrate the physical mechanism responsible for the emergence of time in the classical limit from a fundamentally timeless quantum theory.

In \cite{hajicek:new_PI} a general procedure for passing from the timeless JBB-like theories to time dependent PNM-like theories is given which exploits our relation (\ref{eq:ft}). This makes it possible to define a Hilbert space but at the cost of modifying the Dirac quantization procedure for reparameterization invariant theories. In this paper, we are interested in the emergence of time without having to alter the Dirac procedure.

\subsection{Outline}

The organization of this paper is as follows. We start, in \scn{ham}, by describing the classical equations of motion of JBB theory and develop its Hamiltonian formulation. From this it will be clear that a direct application of Dirac's quantization of gauge systems will lead to a time independent Schr\"odinger equation. This will also be necessary for writing down and gauge fixing the phase space path integral. We will compare JBB theory to a parameterized form of Newtonian mechanics. In contrast to JBB theory, this theory will lead to a time-\emph{dependent} quantum theory. The difference between these two theories will provide a way of understanding the lack of time in JBB theory. In \scn{pi}, we compare the path integrals of each theory and find that we can write one in terms of the other via a Fourier transform. In \scn{time}, we use this result to extract a quantity from the kernel of JBB theory which becomes equivalent to the Newtonian time on-shell. This is our main result which we use to show that there exists a unique notion of duration for quantum systems where the stationary phase approximation is either good or exact.

\section{Hamiltonian Formulation} \label{sec:ham}

Our first task is to write down the canonical form of JBB theory. Other derivations of this exist in the literature and can be found, for example, in \cite{barbour:scale_inv_particles,brown:jacobi_PI,rovelli:bb_qu,smolin:bb_qu,Teitelboim:gravity_proper_t}. We review the main results here since we will need them later and so that we can introduce our notation.

\subsection{Jacobi's Principle and the Timeless Mechanics of Barbour and Bertotti}

\subsubsection{Action and Equations of Motion}

In general, Jacobi's principle is the vanishing of the variation of an action of the form
\equa{
    S = \int d\lambda \sqrt{g_{ab} \dot q^a \dot q^b}
}
where $g_{ab}$ is a metric on a configuration space coordinatized by the $q$'s. This leads to trajectories which are geodesics on configuration space. This form of the action is sufficiently general to include geometrodynamics (after adding an integration over space) and mini-superspace. For our purposes, we are only interested in the global problem of time so it will be sufficient to consider the case where the metric is conformally flat
\equa{\label{eq:metric}
    g_{ab} = -2 V(q) \delta_{ab},
}
where the conformal factor $V(q)$ just corresponds to the potential. It will be convenient to extract the constant part of the potential as this can be interpreted as the negative of the total energy of the system. Thus, we will write $V\ra V - E$ where it is now understood that $V$ should not contain a piece constant in $q$. Because the choice (\ref{eq:metric}) does not affect the structure of the constraints, the problem of time is no different here than it is in mini-superspace so our results regarding time will be generally applicable to symmetry reduced versions of general relativity.

Using the metric (\ref{eq:metric}), the JBB theory is defined by the action

\begin{equation}
\label{eq:sjacobi}
    S_{JBB} = \int_{\lambda_0}^{\lambda_f} d\lambda\quad 2\sqrt{(T(\lambda))(E-V(q_j^i))},
\end{equation}
where $T(\lambda) = \sum_{j=1}^{M} \frac{m_j}{2}\lf(\frac{dq_j^i(\lambda)}{d\lambda}\rt)^2$ is the kinetic energy of an $M$ particle system, $V(q_j^i(\lambda))$ is the potential energy (that does not depend explicitly on $\lambda$) and $E$ is the constant part of $V$ and can be understood as the total energy of the system. The index $i$ ranges from 1 to $d$ while $j$ ranges from 1 to $M$. In this paper we will only consider the case $M=1$ for the sake of compact notation but it is trivial to extend the analysis to the more general case. As can be readily checked, the action (\ref{eq:sjacobi}) is invariant under reparameterizations of $\lambda$ and, as such, its apparent dependence on $\lambda$ is artificial. Thus, $S_{JBB}$ is independent of anything that one could call a time parameter. It does, however, depend on a path $\gamma$ in configuration space. This path is gauge invariant as it represents a collection of points in the configuration space and is independent of any parameter that one might use to parameterize it. Nevertheless, we will artificially introduce $\lambda$, and all the gauge redundancy that goes along with it, so that we can use it as a convenient independent variables in the canonical quantization. This will give us access to well known techniques for determining the correct measure for the path integral.

The classical equations of motion are straightforward to compute. A variation with respect to $q^i$ gives:
\begin{equation}
\label{eq:jacobieom}
    \frac{\sqrt{E - V}}{\sqrt{T}} \frac{d}{d\lambda} \lf( \frac{\sqrt{E - V}}{\sqrt{T}} m\frac{dq^i}{d\lambda}\rt) = -\diby{V}{q^j}\eta^{ij}
\end{equation}
where $\eta^{ij}$ is the flat metric with Euclidean signature. We can then define the reparamaterization invariant quantity
\begin{equation}
\label{eq:timej}
    \tau_{BB} = \int_{\lambda_0}^{\lambda_f} \frac{\sqrt{T}}{\sqrt{E - V}}\, d\lambda
\end{equation}
first referred to as \emph{ephemeris time} by Barbour and Bertotti \cite{barbour:timelessness,barbourbertotti:mach} in analogy to the operational definitions of time first adopted by astronomers \cite{clements:time}. In terms of this quantity, the equations of motion reduce to Newton's equations
\begin{equation}
    m \frac{d^2 q^i}{d\tau_{BB}^2} = -\diby{V}{q^j}\eta^{ij}.
\end{equation}
Thus, $\tau_{BB}$ is equivalent to the Newtonian time but is defined in terms of a length in configuration space equipped with a suitably defined metric\footnote{The metric used to define $\tau_{BB}$ happens to be the \emph{inverse} of the metric used to define the action. Thus, $\tau_{BB}$ is not minimized by the action principle.}. As such, it is a measure of duration that uses the relative change in the positions of the particles in the system and, thus, is a precise realization of Mach's second principle. We will call this system of particles a \emph{Barbour-Bertotti (BB) clock} since it provides us with a way of measuring $\tau_{BB}$.

From the perspective of JBB theory, we start with an action that depends only on the gauge invariant path $\gamma$, which represents the relative positions of particles in the universe, and an arbitrary potential $V(q)$ defined only on configuration space. After writing down the classical equations of motion, we find it convenient to define a gauge invariant quantity called \emph{time} (which can be thought of as a length of $\gamma$) to describe how the $q$'s change relative to each other. In the end, we recover equations of motion equivalent to those of Newton's theory, for a fixed energy $E$, in terms of this invariant quantity. However, in this theory it is not \emph{necessary} to define an absolute Newtonian time: the time \emph{emerges} as a convenient tool for keeping track of the relative positions of particles in a system.

\subsubsection{Hamiltonian Formulation}\label{sec:ham_jbb}

Before quantizing, we must first write down the Hamiltonian formulation of \eq{sjacobi}. To this end, we define the canonical momenta
\equa{
\label{eq:momentaj}
    p_i = \diby{\mathcal{L}_{JBB}}{\dotq^i} = \sqrt{\frac{E-V}{T}} m \dotq^j \eta_{ij}.
}
With this definition, it is easy to see that the canonical Hamiltonian
\begin{equation}
    H_c = p_i \dotq^i - \mathcal{L}(q^i, p_i) = 0
\end{equation}
is identically zero as is always the case in a reparameterization invariant theory.

\eq{momentaj} can be expressed in terms of the quantity $\frac{\dotq^i}{\abs{\dotq^i}}$ and, thus, the momenta $p_i$ should be thought of as unit vectors defining only a \emph{direction} in phase space. As a result, the $p_i$'s don't depend on the length of the $\dotq^i$'s and there is an ambiguity in solving for the $\dotq^i$'s which results in a primary constraint. In this case, the constraint takes the form
\equa{ \label{eq:jacobih} 
	\mathcal{H} = \frac{p^2}{2m} + V(q) - E = 0.
}
It is quadratic in the $p_i$ and can be thought of as a kind of circle identity in phase space. Because it is the only constraint, it is first class. The total Hamiltonian is then proportional to a constraint
\equa{
    H_T = H_c + N(q,p) \mathcal{H} = N(q,p) \mathcal{H},
}
where $N(q,p)$ is an arbitrary Lagrange multiplier.

Next, we define the fundamental Poisson Brackets
\equa{ \label{eq:jacobipb} \{q^i,p_j\} = \delta^i_j}
which we use to compute Hamilton's equations of motion
\begin{align}
    \dotq^i &= \{q^i, H_T\} = N\frac{p_j}{m}\eta^{ij} \label{eq:h1jacobi}\\
    \dot{p}_i &=\{p_i, H_T\} = -N\diby{V}{q^i}.
\end{align}

Before moving on we note that a direct application of Dirac quantization to JBB theory would involve promoting (\ref{eq:jacobih}) to an operator constraint acting on the wavefunction $\Psi(q)$:
\equa{\label{eq:tise}
	\lf[ \frac{\hat{p}^2}{2m} + V(\hat{q}) - E \rt] \Psi(q) = 0.
}
This is simply the time independent Schr\"{o}dinger equation and is the result stated in the introduction (\ref{sec:intro}) that the quantum theory is time independent and leads to a problem of time. Fortunately, the path integral will be more useful in understanding the role of time. Before getting to this however, we will compare JBB theory to another reparameterization invariant theory which, in contrast, leads to a time \emph{dependent} quantum theory.

\subsection{Parameterized Newtonian Mechanics (PNM)}
\label{sec:classicalpnm}

\subsubsection{Action and Equations of Motion}
The reparameterization invariant action of PNM
\equa{
\label{eq:spnm}
    S_{\text{PNM}}(q^i,q^0) = \int_{\lambda_0}^{\lambda_f} d\lambda\, \lf[ \frac{T(\dotq^i(\lambda))}{\dotq^0(\lambda)} - \dotq^0(\lambda) V(q^i(\lambda)) \rt]
}
is defined on \emph{extended configuration space} where $q^0$ is treated as an independent configuration space variable. Classically (and quantum mechanically as we will see), $q^0$ will become the Newtonian absolute time. To see how this happens we vary with respect to $q^i$ giving
\equa{ \label{eq:eompnm}
    \frac{1}{\dotq^0} \frac{d}{d\lambda} \lf( \frac{1}{\dotq^0} m \dotq^i \rt)= -\diby{V}{q^j}\eta^{ij}.
}
These are clearly the Newtonian equations of motion with $t$ replaced by $q^0$. Noting that the action is cyclic in $q^o$ (that is, it only depends on its derivative) a variation with respect to $q^0$ will produce a conserved quantity. This will be the total energy of the system $E$ and the equation of motion for $q^0$ is
\equa{
\label{eq:q0pnm}
    \dotq^0 = \sqrt{\frac{T}{E-V}}.
}
Note the similarities between this theory and JBB theory. If one substitutes \eq{q0pnm} into \eq{eompnm} one gets exactly the equations of motion of JBB theory. Furthermore, \eq{q0pnm} is the definition of BB's ephemeris  time $\tau_{BB}$. In fact, if one adds the boundary term $-E\dotq^0$ to $S_{PNM}$ and substitutes \eq{q0pnm} for $\dotq^0$ one obtains $S_{JBB}$.\footnote{This is a procedure known as the Routhian procedure for eliminating cyclic variables. See \cite{lanczos:mechanics} for further details.} However, there are important differences. Firstly, \eq{q0pnm} is an equation of motion resulting from the variation of an action while \eq{timej} is a simply a \emph{definition}. Secondly, in PNM, $E$ is considered an \emph{integration constant} resulting from integrating the equations of motion and is uniquely determined given a set of boundary conditions for $q^i$ and \eq{eompnm} while, in the JBB theory, $E$ is treated as a \emph{free parameter} of the theory and is used in the definition (\ref{eq:timej}) to uniquely determine $\tau_{BB}$ provided the equations of motion (\ref{eq:jacobieom}) are satisfied. It's as if the roles of energy and time have been switched in terms of how data is inputted into the theory. The connection and differences between these two theories will become very important when trying to see how time might emerge from the path integral quantization of JBB theory. In fact, we will see that the kernels of each theory are related by Fourier transform. The roles of energy and time are particularly important in regards to GR since the quantity analogous to the energy is the cosmological constant. 

\subsubsection{Hamiltonian Formulation} \label{sec:ham_pnm}
We will now write down the Hamiltonian for this system. The canonical momenta are
\begin{align}
    p_i & = \diby{\mathcal{L}_J}{\dotq^i} = \frac{1}{\dotq^0} m \dotq^j \eta_{ij} \notag \\
    p_0 &= \diby{\mathcal{L}_J}{\dotq^0} = -\lf[ \frac{T}{(\dotq^0)^2} + V\rt] \equiv -E.
\end{align}
The primary constraint
\equa{\label{eq:pnmh}
    \mathcal{H} = \frac{p_i^2}{2m} + p_0 + V(q) = 0
}
is a modification of the circle identity for the $p_i$ in the presence of $p_0$. The total Hamiltonian, $H_T = N(q,p) \mathcal{H}$, is a pure constraint. With the fundamental Poisson Brackets $\{q^\alpha, p_\beta\} = \delta^\alpha_\beta$, Hamilton's equations of motion are:
\begin{align}
    \dotq^i &= \{q^i, H_T\} = N\frac{p_j}{m}\eta^{ij} \label{eq:h1pnm}\\
    \dotq^0 &= \{q^0, H_T\} = N \notag \\
    \dot{p}_i &=\{p_i, H_T\} = -N \diby{V}{q^i} \label{eq:h2pnm}\\
    \dot{p}_0 &=\{p_0, H_T\} = 0. \notag
\end{align}
The $\dotp_0$ equation implies conservation of energy. Note the striking similarities between these equations and Hamilton's equations for JBB theory. In both theories, the gauge is fixed by specifying a the function $N$. From the $\dotq^0$ equation of motion, it is clear that $N$ is just the time gauge which parameterizes the path $\gamma$. The simplest gauge that can satisfy the boundary conditions is $\dot N = 0$. In this special gauge, the equations of motion are manifestly equivalent to Newton's equations. Note that, in GR, this is just the \emph{proper time gauge} found to be very useful in \cite{Teitelboim:gravity_proper_t,Teitelboim:quantum_grav}. Note also that our choice of notation for the Lagrange multiplier $N$ is not accidental. The analogous quantity in GR is the lapse which, other than a spacial dependence over the 3-manifold, plays an identical role to that of the $N$ treated here.

\section{On Gauge Invariance and the Phase Space Path Integral}
\label{sec:canqu}

The presence of the primary first class constraint $\mathcal{H}$ in both JBB theory and PNM would normally indicate that we have gauge theories according to the language of Dirac \cite{dirac:lectures}. However, it was argued by Barbour and Foster in \cite{barbour_foster:dirac_thm} that $\mathcal{H}$ should not be thought of as the generator of transformations that leave the physical state of the system unchanged. This has also been noticed by Kucha\v{r} in  \cite{Kuchar:can_qu_gra} and has important implications in regards to the definition of observable quantities in reparameterization invariant theories.

The confusion lies in what one considers the physical state of the system. If one considers a physical state as a point in phase space, as is typically done in the Hamiltonian framework, then $\mathcal{H}$ clearly generates physically \emph{distinguishable} states. To see this, consider the infinitesimal gauge transformations generated by $\mathcal{H}$ in JBB theory. Under $\mathcal{H}$, $q^i$ transforms as
\begin{align}
  q^i(\lambda) &\ra q^i(\lambda) + \epsilon(\lambda) \pb{q^i}{\mathcal{H}} \notag \\
               &= q^i(\lambda) + \epsilon(\lambda) \frac{p_i(\lambda) \eta^{ij}}{m}, \label{eq:H_gen}
\end{align}
which, in general, is clearly a distinct point in phase space. This point is discussed in detail in \cite{Pons:gr_observables} where gauge independent observables are constructed for general relativity and general reparameterization invariant theories.

If, on the other hand, one were to take the view that complete histories, or paths $\gamma$ on configuration space, represent the physical state of a system then $\mathcal{H}$ does indeed generate physically equivalent states. This can be seen by considering the action of $\mathcal{H}$ on a full history. For simplicity and because we will use the notation later, lets consider putting the configuration space on a discrete lattice so that $\lambda$ takes discrete values $\lambda_K$ where capital roman indices range from $1$ to some large number $N$. We then define $\vec{q}_K= \vec{q}(\lambda_K)$ and $\mathcal{H}_K = \mathcal{H}(\lambda_K)$. Using the discrete form of \eq{H_gen}
\equa{
   q_M^i \ra q_M^i + \epsilon_M \frac{p_M^i}{m}
}
and Hamilton's first equation, $H.1.J$, we find that, in a gauge where $N_K = \Delta\lambda_K\, \epsilon_K$,
\begin{equation}
    q_M^i \ra q_M^i + \Delta\lambda_M\,\dot{q}_M^i = q_{M+1}^i.
\end{equation}
Hence, provided $H.1.J$ is satisfied the set $\{\vec{q}_K\}\rightarrow\{\vec{q}_{K+1}\}$. Since this is just a relabeling of the set, we see that $\mathcal{H}$ generates reparameterizations of the history which are, indeed, physically indistinguishable.

In this work we will be treating the path integral for JBB theory and PNM. Though one should be careful to call these theories standard \emph{gauge theories} in light of the previous discussion and the arguments given in \cite{barbour_foster:dirac_thm,Kuchar:can_qu_gra}, as far as the path integral is concerned, it is still valid to use standard gauge theory techniques to determine the measure. The reason for this is that, although the physical states of the path integral are still the in and out states of the kernel (which are configuration space points), when computing the path integral itself, one must sum over histories. Thus, the reparameterization invariance can be treated as a standard gauge redundancy since it will send the complete history to a physically equivalent one. We will then be able to use the standard Faddeev-Popov trick for computing the measure.

Our last task before doing a path integral quantization of JBB and comparing it to PNM is to motivate the use of the phase space path integral. Since the action (\ref{eq:sjacobi}) of JBB is only artificially dependent on $\lambda$ one might wonder if it would be simplest to start with a configuration space path integral with the $\lambda$ dependence removed. In this case, one would no longer have to consider a gauge theory and all the technical complications that come along with it. Unfortunately, the configuration space path integral provides no simple method for determining the measure. In most cases, one must solve for the infinitesimal kernel or integrate the kernel exactly in order to solve for the measure. In the former, one obtains no more information than in the canonical quantization. In the latter, the integration is difficult because of the square roots in the exponential. Moreover, we will find that the measure (given by \eq{fpjbb}) is non-trivial making a comparison to PNM difficult except at the level of the phase space path integral. For these reasons, we find it convenient to keep the $\lambda$ dependence so that we can define momenta and compute the phase space path integral where the precise definition of the measure is understood.

\section{Path Integral Quantization}\label{sec:pi}

We will now use the techniques developed by Faddeev and Popov \cite{faddeev:fp} for gauge fixing the path integral of JBB theory and PNM. We will start with PNM since we can check our results against those of \cite{hartlekuchar:pnm} who present a similar, but not as general, treatment.


\subsection{PNM}

We define the kernel $\kpnmof$ as the phase space path integral (in units where $\hbar = 1$)
\equa{\label{eq:exactpnm}
    \kpnmof = \int \mathcal{D}q^{*\alpha}\, \mathcal{D}p_\alpha^*\, e^{i\int d\lambda\, \lf( p^*_\alpha\dotq^{*\alpha} - \mathcal{H}(q^{*\alpha},p^*_\alpha)\rt)}
}
which is a function of the two configuration space points $q^{\prime\prime\alpha}$ and $q^{\prime\alpha}$. The integration is understood to be over the \emph{true} degrees of freedom $q^{*\alpha}$ and $p_\alpha^*$. Since the true degrees of freedom are, in practice, difficult to solve for explicitly, we would like to write the theory in terms of the redundant variables $q^\alpha$ and $p_\alpha$ then find a gauge fixing condition $\mathcal{G}$ for the first class constraint $\mathcal{H}$. To evaluate the path integral explicitly and to be rigorous about the boundary conditions, we will work with discrete values of $\lambda$ and use the same conventions as \scn{canqu}. This means we should expect the gauge fixing conditions $\mathcal{G}_K$, with Lagrange multipliers $\leps^K$, and the first class constraints $\mathcal{H}^K$, with Lagrange multipliers $N_K$. By inspecting Hamilton's equations for PNM (\ref{eq:h1pnm}) we see that natural gauge fixing conditions are
\equa{\label{eq:pnmgf}
    \mathcal{G}_K = f_K(q^i_K, p_i^K) - \dotq^0_K = 0.
}
In general, the functions $f_K(q^i_K, p_i^K)$ can be nearly arbitrary functions on phase space\footnote{If one is worried about the presence of the $\dotq^0_K$ in the gauge fixing functions recall that we have access to the full history given by the set $\{q_K^{\alpha}\}$ so that we can simply use the definition $\dotq^0_K = \frac{q^0_{K+1}-q^0_K}{\Delta \lambda_K}$ which \emph{is} a function of phase space.} with the only restriction being that they must give a unique solution for $\dotq^0_k$. This has been easily achieved by requiring that the $f_K$ do not depend on the $q^0_K$ or on their conjugate momenta $p_0^K$.

To get the kernel in terms of the partially redundant variables $q^\alpha$ and $p_\alpha$, we must complete $N$ insertions of the identity
\equa{
    1 = \int dG_K\, dH^K\, \delta(\mathcal{G}_K)\, \delta(\mathcal{H}^K)
}
where the constraints are functions of the variables $q^\alpha$ and $p_\alpha$. Making a change of variables from the set $(q^{*\alpha},p^*_\alpha, G,H)\rightarrow (q^\alpha,p_\alpha)$ we pick up a Jacobian factor which is more commonly known as the Faddeev-Popov determinant. With the gauge fixing conditions (\ref{eq:pnmgf}) and the first class constraints (\ref{eq:pnmh}), we get a factor of 1 from commuting the $q^0$'s of the $G_K$ with the $p_0$'s of the $H^K$. Formally we are left with
\equa{
    \fp_{\text{PNM}} = \abs{\pb{f_M(q^i_K, p_i^K)}{\frac{\vecp_N^{\, 2}}{2m}+V^N}}.
}
It is easiest to work out this expression explicitly in specific gauges and for specific choices of $V$. Given these considerations and using the Fourier transform definition of the delta functions, we find
\begin{multline} \label{eq:longkpnm}
    \kpnmof = \int_{-\infty}^\infty \frac{dp_0^0}{2\pi} \frac{d^3\vecp_0}{2\pi} \frac{\Delta\lambda_0 dN_0}{2\pi} \prod_{K=1}^{N-1} \frac{dp_0^K}{2\pi} \frac{d^3\vecp_K}{2\pi} \frac{\Delta\lambda_K dN_K}{2\pi} dq_K^0 \, d^3\vecq_K \, \frac{d\leps^K}{2\pi} \fp_{\text{PNM}}\\
    \times \exp\lf\{ i\sum_{J=0}^{N-1} \Delta\lambda_J \lf[ p_\alpha^J \dotq^\alpha_J - N_J\lf( \frac{\vecp_J^2}{2m}+p_0^J + V^J \rt) - \leps^J\lf( f_J - \dotq^0_J \rt)\rt] \rt\}.
\end{multline}

\subsubsection{Boundary Conditions} \label{sec:pnmbc}

A word or two about boundary conditions are in order. The integrations in \eq{exactpnm} are over $N$ $p$'s but only $(N-1)$ $q$'s as is usually the case. This is, of course, because the boundary conditions impose a constraint on the $q$'s. Similarly, although we need $N$ gauge fixing conditions to make the constraint algebra second class, the boundary conditions impose a constraint on the functions $f_K$. In this case they must satisfy,
\equa{\label{eq:constraint}
    \sum_{J=0}^{N-1} f_J = q^{\prime\prime 0} - q^{\prime 0} \equiv \tau
}
reducing the number of independent gauge fixing functions $f_K$ to $(N-1)$. One can think of the path integral (\ref{eq:longkpnm}) as an integration over $q^0_0$ and $\leps^0$ with the result of imposing the boundary conditions. As a result, we can remove the integrals over $dq^0_0$ and $d\leps^0$ while keeping the constraint algebra second class provided we evaluate the result at $q_0^0 = q^{\prime 0}$ and $q_0^N = q^{\prime\prime 0}$. In this way of thinking, we allow $q_0^0$ to vary but choose gauge fixing functions $f_K$ that guarantee the boundary conditions are satisfied. Thus, it is understood that in the sum of \eq{longkpnm} we should take $\leps^0 = 0$.

\subsubsection{Connection to Standard Quantum Mechanics}

It is possible to connect to the path integral of Hartle and Kucha$\check{\text{r}}$ \cite{hartlekuchar:pnm} using the the proper time gauge discussed in \scn{ham_pnm}. To realize this gauge, we choose functions $f_K = \dot{t}_K$ which are constants over phase space. The Faddeev-Popov determinant is easily seen to be 1. Integrating over the $\leps^K$ gives the infinite product of $\delta$-functions
\equa{
    \prod_{K = 1}^{N-1} \delta(\dotq^0_K - \dot{t}_K)
}
or, equivalently,\footnote{In general the $t$'s can be shifted by the same constant $a$ making the translational invariance of $t$ in standard quantum theory manifest.}
\equa{
    \prod_{K = 1}^{N-1} \delta(q^0_K - t_K).
}
With this, \eq{longkpnm} becomes

\begin{multline}
    \kpnm = \int_{-\infty}^\infty \frac{dp_0^0}{2\pi} \frac{d^3\vecp_0}{2\pi} \frac{\Delta\lambda_0 dN_0}{2\pi} \prod_{K=1}^{N-1} \frac{dp_0^K}{2\pi} \frac{d^3\vecp_K}{2\pi} \frac{\Delta\lambda_K dN_K}{2\pi} dq_K^0 \, d^3\vecq_K \, \delta(q^0_J - t_J)\\
    \times \exp\lf\{ i\sum_{J=0}^{N-1} \Delta\lambda_J \lf[ p_\alpha^J \dotq^\alpha_J - N_J\lf( \frac{\vecp_J^{\,2}}{2m}+p_0^J + V^J \rt) \rt] \rt\}.
\end{multline}
This is exactly the form of $\kpnm$ in \cite{hartlekuchar:pnm}. This confirms that our method does indeed recover standard quantum theory. It is just a special case of the more general kernel given by \eq{longkpnm} which has been obtained using gauge theory techniques. We make the observation that, unlike in \cite{hartlekuchar:pnm}, \eq{longkpnm} allows for more general functions, $f_K$, of phase space provided one computes the correct Faddeev-Popov determinant. Thus, our method allows for more complicated gauge choices.

\subsubsection{$k_{\text{PNM}}$ and Energy Eigenstates}
Before leaving PNM I would like to rewrite \eq{longkpnm} in a form that will allow us to make a connection to JBB theory. First, integrate over the $dp^K_0$'s for \dto{K}{N-1}. This gives the $(N-1)$ $\delta$-functions $\delta(\dotq^0_K - N_K)$. After doing a change of variables from $dq^0_K$ to $d\dot{q}^0_K$ with the definition $\dot{q}^0_K = \frac{q^0_{K+1} - q^0_K}{\Delta\lambda_K}$, an integration over the $d\dot{q}_K^0$ sets $\dotq^0_K = N_K$. The 0-term should be treated separately since we can't integrate over $dq_0^0$. We use the fact that $\Delta\lambda_0 \dotq_0^0 = \tau - \sum_{J=1}^{N-1} \Delta\lambda_0 \dotq_J^0$ and call $p_0^0 \equiv -E$ to write the final integral. Putting this all together gives

\equa{\label{eq:ft}
    \kpnmof = \kpnm(\vecq^{\,\prime\prime},\vecq^{\,\prime},\tau) = \int \frac{dE}{2\pi}\, e^{iE\tau}\, \kpnmt(\vecq^{\,\prime\prime},\vecq^{\,\prime},E)
}
where,

\begin{multline}
    \kpnmt(\vecq^{\,\prime\prime},\vecq^{\,\prime},E) = \int_{-\infty}^\infty \frac{d^3\vecp_0}{2\pi} \frac{\Delta\lambda_0 dN_0}{2\pi} \prod_{K=1}^{N-1} \frac{d^3\vecp_K}{2\pi} \frac{\Delta\lambda_K dN_K}{2\pi} \, d^3\vecq_K \frac{d\leps^K}{2\pi} \fp_{\text{PNM}}\\
    \times \exp\lf\{ i\sum_{J=0}^{N-1} \Delta\lambda_J \lf[ \vecp_J \cdot \dot\vec{q}_J - N_J\lf( \frac{\vecp_J^{\,2}}{2m}-E + V^J \rt) - \leps^J\lf( f_J - N_J \rt)\rt] \rt\}.
\end{multline}
Since we know $\kpnm$ is the kernel for standard quantum mechanics from the results of \cite{hartlekuchar:pnm}, $\int d\tau\, e^{i\tau E}\, \kpnm$ will give the kernel for energy eigenstates of energy $E$. One immediately recognizes $\kpnmt(E)$ as this kernel.

\subsection{JBB Theory} \label{sec:jacobi}

We now consider the path integral of JBB theory. It is worth noting that this path integral has been considered before in different settings. One can find the BRST quantization of JBB theory in references \cite{Teitelboim:gravity_proper_t} and \cite{brown:gr_time}. Also, for the BRST treatment of the relativistic particle (which is nearly mathematically identical to JBB) see \cite{Henneaux_tiet:susy_part}. However, while all of these papers use ghost fields to do the gauge the fixing, we use the Faddeev-Popov trick directly so that we can be completely rigorous about boundary conditions. This form allows us to carefully compare JBB with PNM so that we can extract time explicitly from the JBB kernel.

Making use of the same techniques used to write the phase space path integral for PNM, we choose the gauge fixing functions
\equa{
    \mathcal{G_K} = f_K(q^i_K, p_i^K) - \frac{m\vecp_K\cdot\dot\vec{q}_K}{p_K^2}=0.
}
The phase space path integral is then
\begin{multline}
\label{eq:kj}
    \kj (\vecq^{\,\prime\prime},\vecq^{\,\prime},E) = \int_{-\infty}^\infty \frac{d^3\vecp_0}{2\pi} \frac{\Delta\lambda_0 dN_0}{2\pi} \prod_{K=1}^{N-1} \frac{d^3\vecp_K}{2\pi} \frac{\Delta\lambda_K dN_K}{2\pi} \, d^3\vecq_K \frac{d\leps^K}{2\pi} \fp_{JBB}\\
    \times \exp\lf\{ i\sum_{J=0}^{N-1} \Delta\lambda_J \lf[ \vecp_J \cdot \dot\vec{q}_J - N_J\lf( \frac{\vecp_J^{\,2}}{2m}-E + V^J \rt) - \leps^J\lf( f_J - \frac{m\vecp_J\cdot\dot\vec{q}_J}{p_J^2} \rt)\rt] \rt\}
\end{multline}
where we impose the boundary conditions by evaluating this at $\vecq_N = \vecq^{\,\prime\prime}$ and $\vecq_0 = \vecq^{\,\prime}$. In the above, as with PNM, it is understood that $\leps^0 = 0$. The Faddeev-Popov determinant is easiest to write out in specific gauges. It can be formally written as
\equa{\label{eq:fpjbb}
    \fp_{\text{JBB}} = \abs{\pb{f_M - \frac{m\vecp_M\cdot\dot\vecq_M}{p_M^2}}{\frac{\vecp^{\,2}_N}{2m}+V^N}}.
}

\subsubsection{Boundary Conditions} \label{sec:bcj}
For \scn{time} we will need to know explicitly how the boundary conditions have been imposed in \eq{kj}. This complication arises because of the fact that, of the three independent components of the vectors $\vecq_K$, one of them is a pure gauge. Thus, two of the boundary conditions can be imposed in the usual way; while, the third condition, like the case of PNM described in \scn{pnmbc}, should be imposed by letting the gauge degree of freedom vary freely and by choosing gauge fixing functions $f_K$ that guarantee that the boundary conditions will be satisfied.

To see this realized explicitly, we must add to \eq{kj} an integration over $d\leps^0$ and $d^3\vecq_0$ and we must include the term $\exp(i\Delta\lambda_0\leps^0(f_0 - \frac{\vecp_0\cdot\dot\vecq_0}{p_0}))$ in the integrand. We will need some notation to split the gauge piece of $\vecq_0$ from the physical piece. For an arbitrary vector $\vec{x}$, we define $x^{||}\equiv \frac{\vec{x}\cdot\vecp_0}{p_0}$, and $x^{\bot} \equiv \frac{\abs{\vec{x}\times\vecp_0}}{p_0}$. These definitions allow us to write any vector in terms of a cylindrical coordinates system where $\vecp_0$ points in the z-direction. The boundary conditions for the non-gauge degrees of freedom can be imposed simply by integrating over $d\phi_0 dq^{\bot}_0$ and $d\phi_N dq^{\bot}_N$ after inserting the $\delta$-functions $\delta(\phi_{q'}-\phi_{0})\delta(\vec{q}^{\,\prime\bot} - q^{\bot}_0)$ and $\delta(\phi_{q''}-\phi_{N})\delta(\vec{q}^{\,\prime\prime\bot} - q^{\bot}_N)$. For the gauge degrees of freedom, it is a short calculation to show that, choosing $f_0$ such that
\equa{\label{eq:jacobiconstraint}
    \frac{m (q^{\prime\prime ||} - q^{\prime ||})}{p_0} + \sum_{J=0}^{N-1} \Delta\lambda_J \lf[ m\dot\vecq_J\cdot \lf( \frac{\vecp_J}{p_J^2} - \frac{\vecp_0}{p_0^2} \rt) - f_J \rt] = 0
}
will lead to the appropriate boundary conditions for $q_0^{||}$. Although this seems like an unnecessary amount of work just to justify the integration over $\leps^0$ and $d^3\vecq_0$, we will see that the difference between a time dependent kernel and a time independent kernel is exactly expressed by the form of the constraint (\ref{eq:jacobiconstraint}). We can now see that the advantage of working with this formalism is that it gives us a precise way to compare the roles of time both theories.

\subsection{Connection Between $k_{\text{PNM}}$ and $k_{\text{JBB}}$}

It is constructive to rewrite $\kpnm(E)$ and $\kj (E)$ in special gauges. For PNM, we pick the gauge
\equa{
    f_K = \frac{m\vecp_K\cdot\dot\vec{q}_K}{p_K^2}.
}
The Faddeev-Popov determinant takes the form
\equa{
    \fp_{\text{PNM}} = \abs{\pb{\frac{m\vecp_M\cdot\dot\vecq_M}{p_M^2}}{V^N}}.
}
For JBB theory, we pick
\equa{
    f_K(q^i_K, p_i^K) = N_K.
}
These conditions lead to the same Faddeev-Popov determinant as PNM. By comparing the integrands, we see that in these gauges it is manifest that $\kpnmt(E) = \kj (E)$. In other words, the two kernels are related by the Fourier transform (\ref{eq:ft}). This agrees with a result derived in \cite{brown:jacobi_PI}. For completeness, we write out the full expression for the kernels in this gauge:
\begin{multline}
     \kpnmt(E) = \kj (E) = \int_{-\infty}^\infty \frac{d^3\vecp_0}{2\pi} \frac{\Delta\lambda_0 dN_0}{2\pi} \prod_{K=1}^{N-1} \frac{d^3\vecp_K}{2\pi} \frac{\Delta\lambda_K dN_K}{2\pi} \, d^3\vecq_K \frac{d\leps^K}{2\pi} \abs{\pb{\frac{m\vecp_K\cdot\dot\vecq_K}{p_K^2}}{V^L}}\\
    \times \exp\lf\{ i\sum_{J=0}^{N-1} \Delta\lambda_J \lf[ \vecp_J \cdot \dot\vec{q}_J - N_J\lf( \frac{\vecp_J^2}{2m}-E + V^J \rt) - \leps^J\lf( N_J - \frac{m\vecp_J\cdot\dot\vec{q}_J}{p_J^2} \rt)\rt] \rt\}.
\end{multline}

Since these kernels are manifestly the same in the gauges described above, they must also be the same in any gauge. Thus, the straightforward path integral quantization of JBB theory gives the kernel for energy eigenstates of energy $E$. We have recovered the standard result obtained by canonical quantization where the solutions are stationary states and time, it seems, has disappeared.

\section{Time and the Stationary Phase Approximation}
\label{sec:time}

It is now possible to see the difference between the time dependent kernel of PNM and the time independent kernel of JBB theory. In \scn{bcj}, we noted that, in order to impose the boundary conditions, we should choose $f_0$ such that \eq{jacobiconstraint} is satisfied. Implementing the procedure outlined in that section we find that we can rewrite $\kj$ as
\begin{multline}\label{eq:jbb_flow_kernel}
    \kj (\vecq^{\,\prime\prime},\vecq^{\,\prime},E) = \int_{-\infty}^\infty \frac{d\leps^0}{2\pi}\lf[ \int_{-\infty}^\infty \frac{d^3\vecp_0}{2\pi} \frac{\Delta\lambda_0 dN_0}{2\pi}\, d^3\vecq_0\,\delta(\phi_{q'}-\phi_{0})\delta(\vec{q}^{\,\prime\bot} - q^{\bot}_0)\rt.\\\lf.
    \times\prod_{K=1}^{N-1} \frac{d^3\vecp_K}{2\pi} \frac{\Delta\lambda_K dN_K}{2\pi} \, d^3\vecq_K \frac{d\leps^K}{2\pi} \fp_{JBB}\rt.\\\lf.
    \times \exp\lf\{ i\sum_{J=0}^{N-1} \Delta\lambda_J \lf[ \vecp_J \cdot \dot\vec{q}_J - N_J\lf( \frac{\vecp_J^{\,2}}{2m}-(E+\leps^0) + V^J \rt) - \leps^J\lf( f_J - \frac{m\vecp_J\cdot\dot\vec{q}_J}{p_J^2} \rt)\rt] \rt\} \rt]\exp(i\leps^0\tau)
\end{multline}
where,
\equa{\label{eq:tau}
    \tau = \frac{m (q^{\prime\prime||} - q^{\prime ||})}{p_0} + \sum_{J=0}^{N-1} m \dot\vecq_J\cdot\lf( \frac{\vecp_J}{p_J^2} - \frac{\vecp_0}{p_0^2} \rt).
}

The bracketed expression after the $d\leps^0$ integral is \emph{nearly} equal to $\kpnmt (E+\leps^0)$. If it were and if we were able to pull the factor $e^{i\leps^0 \tau}$ through the integral in the bracketed expression then we would have
\equa{\label{eq:RG_flow}
    \kj^{\prime}(\vecq^{\,\prime\prime},\vecq^{\,\prime},E) = \int \frac{d\leps^0}{2\pi} e^{i\leps^0\tau} \kpnmt(\vecq^{\,\prime\prime},\vecq^{\,\prime},E+\leps^0).
}
which is $\kpnm(\tau)$ up to an unobservable global $U(1)$ factor $e^{-iE\tau}$. That is, we would have a theory with time. But $e^{i\leps^0 \tau}$ cannot, in general, be moved through the integral since $\tau$ is a complicated function of phase space. Furthermore, the bracketed expression is missing the appropriate boundary condition $\delta$-function that would make it exactly equal to $\kpnmt$. Hence, if we want time to emerge in the quantum JBB theory, we must: a) find a way to implement the boundary conditions separately from putting constraints on the gauge fixing conditions, and b) we must be able to pull $\tau$ through the integral over all of phase space. This is possible in the stationary phase approximation.

In the stationary phase approximation we approximate the kernel by a sum over the unique history that extremizes the action\footnote{It is possible, in more general cases, that more than one solution will extremize the action. In these cases, the emergent time could be different depending on the observed vacuum state of the system.}. That is, we approximate the kernel by a sum over the classical history. Because we no longer have an integral over all of phase space, $\tau$ can be moved through the bracketed expression. Furthermore, the boundary conditions are imposed by requiring the classical solution. Thus, we have succeeded in showing that the stationary phase approximation gives us a theory with time. However, this is not a theory with just any time. The emergent time must be given by \eq{tau} which is a specific function of the classical history. Using the boundary conditions and returning to the continuous limit, we see that $\tau = \tau_{BB}$. That is, the time that is emergent in the stationary phase approximation is exactly the ephemeris time of the classical theory. It is the time read off by a BB-clock.

On one hand, this should not be surprising since the stationary phase approximation should simply be recovering the classical limit. On the other hand, this may seem unexpected in light of the canonical quantization since we are recovering a classical limit where time is well defined even though the quantum theory is governed by the time-independent Schr\"odinger equation. Somehow, we've found that time can ``live'' in the time-independent Schr\"odinger equation. Furthermore, the stationary phase approximation extends much further than just the classical limit since it is exact for up to quadratic potentials.

It is important to note that, like in the classical theory, the roles played by time and energy in the way data is inputted into the stationary kernels of PNM and JBB are switched. That is, one cannot simply plug a time into the kernel of JBB just as one cannot simply plug an energy into the kernel of PNM. However, in the stationary phase approximation, a unique energy can be calculated for a unique time simply by inverting \eq{timej} and inserting the classical history. Thus, the algorithm for comparing the two theories involves either specifying a time $t$ for $\kpnm(\vec{q}'',\vec{q}',t)$ then calculating the energy $E(t)$ by inverting \eq{timej} to insert into $\kj(\vec{q}'',\vec{q}',E)$ or specifying an energy $E$ for $\kj(\vec{q}'',\vec{q}',E)$ then calculating $t(E)$ using \eq{timej} to insert into $\kpnm(\vec{q}'',\vec{q}',t)$. Specifically, we have shown that, in the stationary phase approximation, we have the equality
\equa{ \label{eq:equality} \kpnm(\vec{q}'',\vec{q}',t, E(t))= e^{iE\tau} \kj(\vec{q}'',\vec{q}',t(E), E).}

This agrees with our intuition from the classical theory. On shell, the emergent time is determined through \eq{timej} uniquely by specifying the energy $E$ and by imposing the boundary conditions and the classical equations of motion. Off shell however, \eq{timej} gives a different Barbour-Bertotti time for each history since an arbitrary history will lead, in general, to a very different value of $\tau_{BB}$ for a fixed energy. Because we sum over all histories, each contribution to the kernel will represent a different Barbour-Bertotti clock leading to a kind of superposition of clocks\footnote{Note that it is the action and not the ephemeris time that is actually being summed over in the path integral.}. This superposition effectively integrates time out of the theory and leads to solutions of the quantum theory that are stationary states. From the path integral perspective, the mechanics responsible for this is very clear. The classical theory does have a unique notion of duration because, in the stationary phase approximation, there is only one time that gives an important contribution to the kernel: the Barbour-Bertotti time.

Decoherence provides an alternative perspective for understanding this emergence. Our path integral result suggests that, in the stationary phase approximation, the Barbour-Bertotti time decoheres from the other components of the superposition making it a useful clock. A more detailed study of decoherence in the context of emerging clocks can be found in \cite{Kiefer:semi_class_review}. We only note here that our intuition from the path integral seems to agree with our expectations from decoherence.

\section{Outlook}

We have seen that the path integral quantization of JBB theory leads to a theory whose solutions are energy eigenstates and are thus governed by the time independent Schr\"odinger equation. By comparing this time independent theory to the manifestly time dependent kernel of PNM we were able to see that the timelessness is a result of a superposition of all possible BB-clocks which occurs when one sums over all possible histories. Nevertheless, a unique time can be recovered in the classical limit or, more generally, in the stationary phase approximation because the path integral is dominated by contributions due to the unique classical history. From a technical perspective, the difference between the time dependent PNM theory and the time independent JBB theory is
in the constraints applied to the gauge fixing functions to impose the boundary conditions.

However, we still have a lot of work to do before we can make this notion of time more precise in the quantum theory. For example, we have not yet demonstrated how a wavefunction might evolve unitarily in terms of $\tau_{BB}$ even in situations where the stationary phase approximation is exact. Furthermore, in situations where the stationary phase approximation fails, it seems that this notion of time breaks down completely. Nevertheless, we may not be doomed.

It is well known that (see for instance \cite{Kiefer:semi_class_review}) in the context of gravity coupled to a scalar field that one can use a Born-Oppenheimer (BO) ansatz to show that \emph{heavy} (ie, semi-classical) degrees of freedom can provide a kind of BB-clock under which \emph{light} degrees of freedom evolve according to the time dependent Schr\"odinger equation. These results are very interesting but they rely on use of the BO ansatz which has the slight drawback that it makes assumptions about the emergent theory that may not necessarily be implied by the fundamental theory even though they are consistent with it. One example of such an assumption is the presence of a \emph{complex} wavefunction which is needed in order to make the BO ansatz but not necessary if one is starting from the time-independent Schr\"odinger equation. For more detailed discussions on this particular point see \cite{Barbour:time_and_c_numbers} and the rebuttal in \cite{Kiefer:semi_class_review}.

An alternative option, which is suggested by the results of this paper, would be to treat these toy models and assume only a mass gap between heavy and light degrees of freedom. Then one could construct a Wilsonian effective action which would integrate out the heavy particles. The hope would be that the RG flow towards the IR would turn the JBB action into the PNM action with a specific expression for the emergent time. Equations (\ref{eq:jbb_flow_kernel}) and (\ref{eq:tau}) tell us exactly how one theory should flow into the other and suggest the that energy might play the role of an order parameter. If it works, this ``bottom up'' approach could show how a time dependent Schr\"odinger equation might emerge for light degrees of freedom without having to resort to the BO ansatz and could, for example, shed light on why the complex numbers necessarily arise.


\begin{acknowledgments}
    I would like to extend a special thanks to Hans Westman for introducing me to JBB theory and for many thought provoking discussions. I would also like to thank Rafael Sorkin for pointing me to the paper of Hartle and Kucha$\check{\text{r}}$, Ed Anderson for providing many useful references and suggestions, Julian Barbour for helpful comments and clarifications, and Lee Smolin for invaluable depth of knowledge and guidance. I would also like to thank the referee for helpful suggestions and observations which helped clarify the presentation. Research at the Perimeter Institute is supported in part by the Government of Canada through NSERC and by the Province of Ontario through MEDT. I also acknowledge support from an NSERC Postgraduate Scholarship, Mini-Grant MGA-08-008 from the Foundational Questions Institute (fqxi.org), and from the University of Waterloo.
\end{acknowledgments}




\bibliographystyle{utphys}
\bibliography{mach}

\end{document}